\documentclass[aps,pre,reprint,groupedaddress]{revtex4-1}
\usepackage{amssymb,amsfonts,amsmath}
\usepackage[dvips]{graphicx,color,psfrag}

\begin{document}


\title{
Power law in random multiplicative processes
with spatio-temporal correlated multipliers
}


\author{Satoru Morita}
\email[]{morita.satoru@shizuoka.ac.jp}
\affiliation{Department of Mathematical and Systems Engineering, Shizuoka University, Hamamatsu, 432-8561, Japan}

\date{\today}

\begin{abstract}
It is well known that random multiplicative processes
generate power-law probability distributions. 
We study how the
spatio-temporal correlation of the multipliers
influences the power-law exponent.
We investigate two sources of the time correlation: the local environment and the global environment.
In addition, we introduce two simple models through which we analytically and numerically show that the local and global environments yield different trends in the power-law exponent.
\end{abstract}

\pacs{05.40.-a, 02.50.-r, 87.10+e, 89.65.Gh}

\maketitle

Power-law distributions are ubiquitous not only in 
natural systems but also in social systems \cite{newman05,clauset,gabaix09}.
For instance, 
city sizes \cite{gabaix99,ioannides}, 
firm sizes \cite{ramsden,axtell}, 
stock returns \cite{mandelbrot,gabaix03}, and
personal incomes \cite{champernowne,reed}
follow the power-law 
\begin{equation}
P(x)\propto x^{-\gamma-1}.
\label{eq1}
\end{equation}
over large scales. This expression is widely known as Pareto's law \cite{pareto} or 
Zipf's law \cite{zipf}, 
and it has been well investigated using various models.
One well-known mechanism that generates power-law distribution is the random multiplicative process (RMP) \cite{render,sornette,biham,sornette98,nakao,sato}. 

This paper aims
to clarify the influence of the 
spatio-temporal correlation of the multipliers on the power-law exponent $\gamma$.
$\gamma$ is known to decrease as the correlation time length increases \cite{sato}.
However, there is no general formulation of RMPs with
spatio-temporal correlation.
To fulfill our aim, we consider two simple models,
Model 1 and Model 2, in which the temporal correlation is led by the local environment and the global environment, respectively.
We show analytically and numerically that
the correlation time length influences
the power-law exponent $\gamma$ in Model 1, whereas
it does not in Model 2.

This paper is organized as follows.
First, we introduce 
a binomial multiplicative process, 
in which the random multipliers can have only two values,
and revisit the case where the multipliers have no correlation.
Here, we propose a graphical method to estimate the exponent 
$\gamma$.
Next, we separately analyze the cases of temporal and spatial correlation.
We then analyze
the effect of spatio-temporal correlation in Models 1 and 2.
The results of the numerical simulations are compared with theoretical predictions.
The paper concludes with a summary.

Here, we consider a simple version of RMP,
\begin{equation}
x_i(t+1)=m_i(t)x_i(t)+b,
\label{eq2}
\end{equation}
where $t$ denotes a discrete time step and
$i$ specifies the elements ($i=1,2,3,\dots,N$).
The variable $x_i(t)$ can represent any quantity such as population,
firm size, or income,
$m_i(t)$ is a random multiplier, and 
$b$ is an additive positive term.	
Although $b$ may be a stochastic variable,
we set $b$ as a constant because it does not influence 
the following results. 
The initial condition is set to $x_i(0)=1$ so that 
$x_i(t)$ is always positive.
For simplicity, we assume that the stochastic multiplier
$m_i(t)$ is $m_+$ or $m_-$, each with probability $1/2$.
Here, we set $m_+>1>m_->0$
and $\ln m_+ + \ln m_- <0$, so that 
eq.~(\ref{eq2}) has a stationary distribution.
The cases of $m_+$ and $m_-$ are called good and bad
environments, respectively.
The mean of the multipliers is $\mu=(m_++m_-)/2$, and
their variance is given by $\sigma^2=(m_+-m_-)^2/4$.

First, we consider 
the case of no correlation between the multipliers
$m_i(t)$:
\[
\langle (m_i(t)-\mu)(m_j(t')-\mu) \rangle=\sigma^2\delta_{ij} \delta_{tt'}.
\]
The stationary distribution has a power-law tail,
where the exponent $\gamma$ is 
a positive root of
\begin{equation}
\langle m^{\gamma} \rangle
=\left({m_+}^{\gamma}+{m_-}^{\gamma}\right)/2=1.
\label{eq3}
\end{equation}
The power-law tail 
exists because $\ln x_i(t)$ undergoes a random walk with a drift
toward smaller values and is repelled from $-\infty$ 
\cite{sornette,biham}. 
Consequently,
$\ln x_i(t)$ follows an exponential distribution,
implying a power-law distribution in $x_i(t)$. 
Condition (\ref{eq3}) was proved 
by Kesten \cite{kesten} using renewal theory.
Intuitively, Eq. (\ref{eq3}) can be derived as follows \cite{sato}.
In the large-scale range ($x \gg b$),
we raise both sides of eq. (\ref{eq2}) to the power $\alpha$ 
and average them. Thus, we have
\[
\langle x_i(t)^{\alpha}\rangle \simeq 
\langle m^{\alpha}\rangle^t.
\]
If eq.~(\ref{eq1}) holds, 
$\langle x_i(t)^{\alpha}\rangle$ will converge
when $\alpha<\gamma$. 
Consequently, the exponent $\gamma$ is given by the critical value of $\alpha$, 
i.e., by eq.~(\ref{eq3}).

In many systems, $\gamma$ is approximately one \cite{newman05,clauset}.
In our model, $\gamma$ is $1$ when $(m_++m_-)/2=1$.
Thus, if the values of $m_+$ and $m_-$ lie along the solid curve in Fig.~1, 
then $\gamma=1$. Note that the condition is a
function of ${m_+}^{\gamma}$ and ${m_-}^{\gamma}$.
Therefore, from the plot in Fig.~1, we can graphically estimate the exponent $\gamma$ for general values of $m_+$ and $m_-$.
When the parameters $m_+$ and $m_-$ are located beneath the solid curve, 
$\gamma$ is larger than 1.
In the parameter region between this curve and the diagonal dotted lines, $\gamma<1$.
\begin{figure}[tb]
\includegraphics[width=0.36\textwidth]{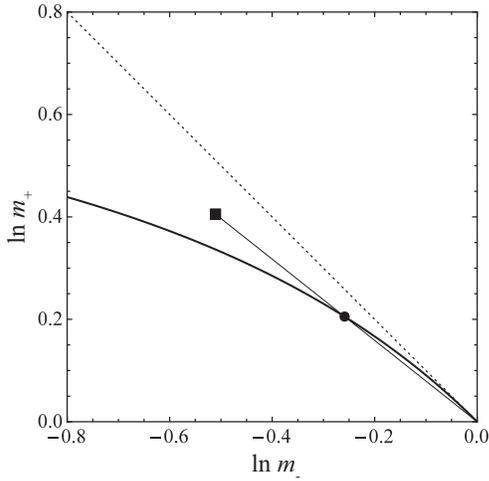}
\caption{
Graphical method for estimating the exponent $\gamma$.
The vertical axis is $\ln m_+$ and 
the horizontal axis is $\ln m_-$.
The solid curve represents $(e^{\ln m_+}+e^{\ln m_-})/2=1$,
i.e., $\gamma=1$.
For general values of $m_+$ and $m_-$, $\gamma$ 
satisfies $(e^{\gamma\ln m_+}+e^{\gamma\ln m_-})/2=1$; thus,
 $\gamma$ can be estimated using this curve.
For example, consider the case of $m_+=1.5$ and $m_-=0.6$
(indicated by the closed square).
Draw a straight line between the square and the origin $(0,0)$.
The closed circle represents the intersection of the curve and the straight line.
The exponent $\gamma$ is determined by 
the ratio of the distance between the circle and the origin 
to the distance between the square and the origin.
In this case, we obtain $\gamma=0.506\dots$.
Above the diagonal dotted line ($\ln m_+ + \ln m_->0$),
there is no stationary distribution.}
\label{fig1}
\end{figure}

\begin{figure}[tbh]
\includegraphics[width=0.36\textwidth]{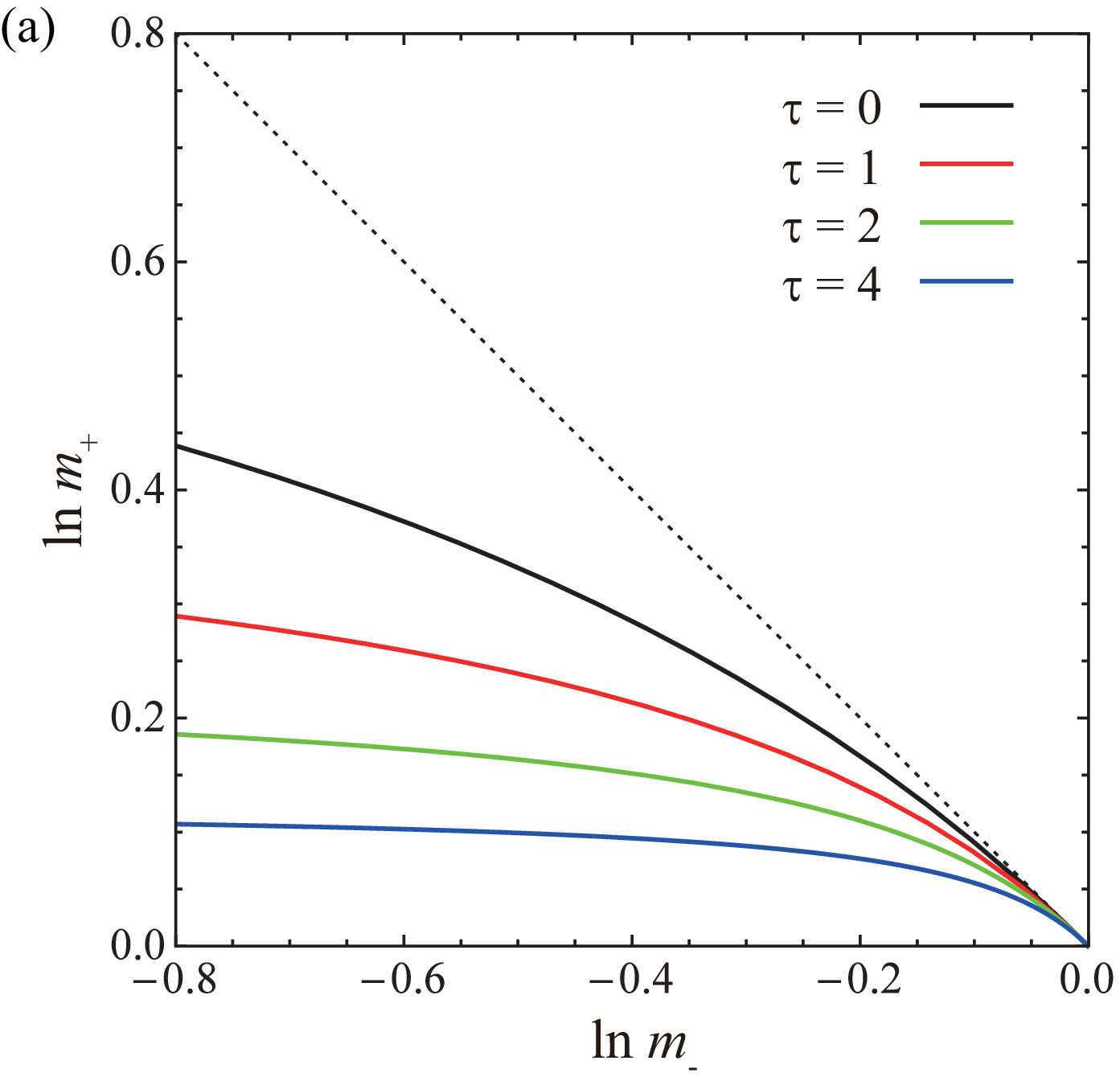}
\includegraphics[width=0.36\textwidth]{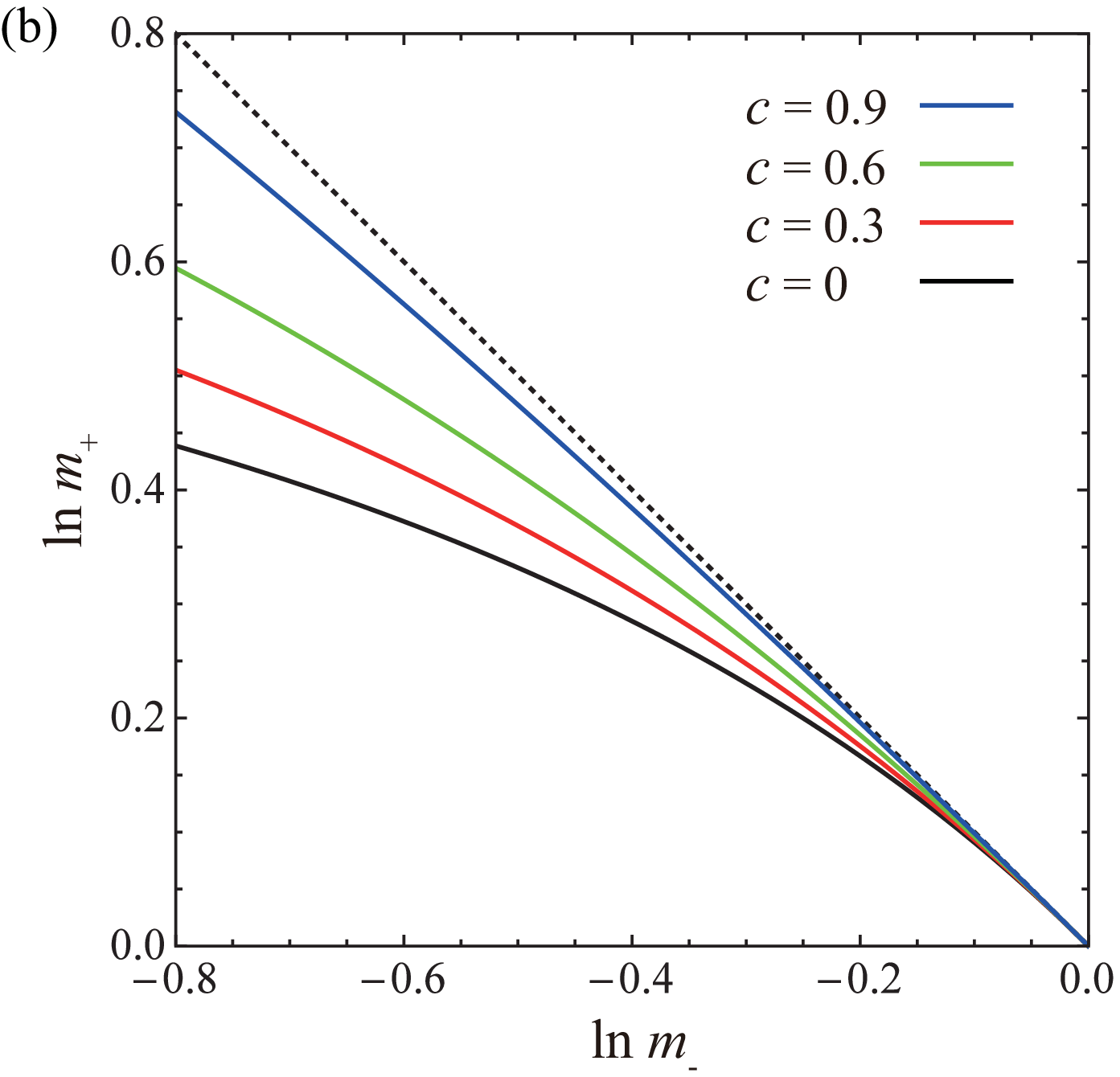}
\caption{
A set of $m_+$ and $m_-$ for which $\gamma=1$ for several cases. From these curves, we can estimate the exponent $\gamma$ 
for general values of $m_+$ and $m_-$, as in Fig.~1.
(a)
The case in which $m_i(t)$ has only temporal correlation (\ref{eq4}).
The correlation time length is set as $\tau=0, 1, 2, 4$.
The curves are given by 
$2-{m_+}- {m_-} + e^{-1/\tau}(2{m_+}{m_-}
 - {m_+}-{m_-}) =0$.
(b)
The case in which $m_i(t)$ has only spatial correlation (\ref{eq7}).
The correlation coefficient is set as $c=(1-2p)^2=0, 0.3, 0.6, 0.9$.
The curves represent 
$(p{m_+}+q{m_-})(p{m_-}+q{m_+})=1$.}
\label{fig2}
\end{figure}
Second, we consider the case of temporally correlated $m_i(t)$:
\begin{equation}
\langle (m_i(t)-\mu)(m_j(t')-\mu) \rangle
=\sigma^2\delta_{ij} e^{-|t-t'|/\tau},
\label{eq4}
\end{equation}
where $\tau$ is the correlation time length.
To obtain a stochastic time series satisfying 
(\ref{eq4}), we use a Markov
chain with the transition probability matrix
\begin{equation}
A=\left(
\begin{array}{cc}
1-u/2 & u/2\\
u/2 & 1-u/2
\end{array}
\right),
\label{eq5}
\end{equation}
where $u=1-e^{-1/\tau}$.
For example, the first-row, second-column element of the matrix
$A$ represents the probability that the multiplier changes from
$m_-$ to $m_+$.
Let $x_{+}(t;\alpha)$ and $x_{-}(t;\alpha)$ be defined as 
the total $x_i(t)^{\alpha}$ in the good and bad local environments,
respectively: 
\[
x_{\pm}(t;\alpha)=\sum_{i\in \{i|m_i(t)=m_{\pm}\}} {x_i}^{\alpha}(t).
\]
Then, 
\[
\left(
\begin{array}{c}
x_+(t+1;\alpha)\\x_-(t+1;\alpha)
\end{array}
\right)\simeq A
\left(\begin{array}{cc}
{m_+}^{\alpha}&0\\
0&{m_-}^{\alpha}
\end{array}
\right)
\left(
\begin{array}{c}
x_+(t;\alpha)\\x_-(t;\alpha)
\end{array}
\right).
\]
Since the exponent $\gamma$ is the critical value of $\alpha$, 
it is determined by solving the following equation such that 
the dominant eigenvalue of 
\[
A
\left(\begin{array}{cc}
{m_+}^{\gamma}&0\\
0&{m_-}^{\gamma}
\end{array}
\right)
\]
equals one.
Using a simple calculation, we show that
the exponent $\gamma$ is given by a solution of 
\begin{equation}
2-{m_+}^{\gamma} - {m_-}^{\gamma} + e^{-1/\tau}(2{m_+}^{\gamma}{m_-}^{\gamma}
 - {m_+}^{\gamma} -{m_-}^{\gamma}) =0 .
\label{eq6}
\end{equation}
Figure 2(a) plots the curves of (\ref{eq6}) for $\gamma=1$ 
for various values of $\tau$.
For general values of $m_+$ and $m_-$,
the exponent $\gamma$ can be estimated from these curves,
as demonstrated in Fig.~1.
This result indicates that the exponent
$\gamma$ decreases as the correlation time length increases,
which is consistent with the previous work by Sato et al.~\cite{sato}.

Third, we consider the case of spatially correlated $m_i(t)$:
\begin{equation}
\langle (m_i(t)-\mu)(m_j(t')-\mu) \rangle
=\sigma^2\delta_{tt'} c
\label{eq7}
\end{equation}
where $i\neq j$ and $c$ is the correlation coefficient.
We now introduce the global environment, which can independently be
good or bad with probability $1/2$ at each time.
In a good global environment at time $t$,
the multiplier $m_i(t)$ is $m_+$ with 
probability $p>1/2$
or $m_-$ with probability $q=1-p$.
Conversely, in a bad global environment, $m_i(t)$ is $m_-$ with 
probability $p$ or $m_+$ with probability $q$.
By simple algebra, we obtain
\[
c=(1-2p)^2.
\]
When $t\gg 1$,
the mean of $x_i(t)^{\alpha}$ is given by 
the geometrical mean of the average growth rates 
in both global environments:
\[
\langle x_i(t)^{\alpha}\rangle\simeq
(p{m_+}^{\alpha}+q{m_-}^{\alpha})^{t/2}
(p{m_-}^{\alpha}+q{m_+}^{\alpha})^{t/2}.
\]
Thus, the exponent $\gamma$ is obtained by solving 
\begin{equation}
(p{m_+}^{\gamma}+q{m_-}^{\gamma})
(p{m_-}^{\gamma}+q{m_+}^{\gamma})=1.
\label{eq8}
\end{equation}
Figure 2(b) plots the curves (\ref{eq8}) for $\gamma=1$ for various spatial correlations $c$.
The exponent $\gamma$ is graphically estimated as before.
We find that $\gamma$ increases with increasing spatial correlation.

To analyze the spatio-temporal correlation of $m_i(t)$, we expand the previous model and introduce two new models,
Model 1 and Model 2, in which the time correlation is sourced from the local and global environments, respectively.
In Model 1, the local environment $m_i(t+1)$ in a good global environment is governed by the transition probability matrix 
\[
A_+=\left(
\begin{array}{cc}
1-q u & p u\\
q u & 1-p u
\end{array}
\right),
\]
where $u=1-e^{-1/\tau}$.
In a bad global environment, the transition probability matrix is 
\[
A_-=\left(
\begin{array}{cc}
1-p u & q u\\
p u & 1-q u
\end{array}
\right).
\]
The global environment is independently good or bad with 
probability $1/2$; therefore, the transition probability matrix of the multiplier
$m_i(t)$ for each $i$ should be the average of $A_+$ and $A_-$ ($A$), which is given in eq. ($\ref{eq5}$).
Thus, the autocorrelation of $m_i(t)$ is $e^{-|t-t'|/\tau}$.
Moreover, a simple calculation gives 
\begin{equation}
\langle (m_i(t)-\mu)(m_j(t')-\mu) \rangle=\left\{
\begin{array}{lcl}
\sigma^2e^{-|t-t'|/\tau} &\ & (i= j)\\
\sigma^2 c \ e^{-|t-t'|/\tau} &\ & (i\neq j),
\end{array}
\right.
\label{eq9}
\end{equation}
where the correlation coefficient $c$ is given by 
\[
c=(1-2p)^2\frac{u}{2-u}.
\]

\begin{figure}[tb]
\includegraphics[width=0.48\textwidth]{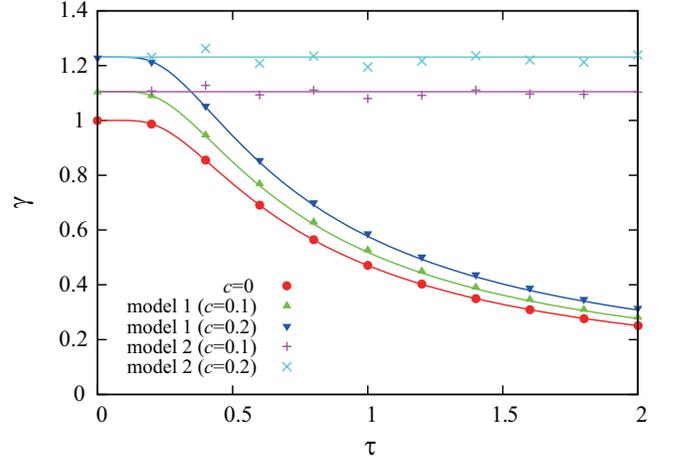}
\caption{Exponent $\gamma$ of the power-law tail
as a function of the correlation time $\tau$.
Here, we estimate $\gamma$ for $N=10^6$ and 200 samples.
The other parameters are set to $m_+=1.5$, $m_-=0.5$.
The symbols and curves represent the numerical estimations and theoretical predictions, respectively.
}
\label{fig3}
\end{figure}

In this case, the dynamics of $x_{\pm}(t;\alpha)$ are determined as
\[
\left(
\begin{array}{c}
x_+(t+1;\alpha)\\x_-(t+1;\alpha)
\end{array}
\right)=
A_+
\left(\begin{array}{cc}
{m_+}^{\alpha}&0\\
0&{m_-}^{\alpha}
\end{array}
\right)
\left(
\begin{array}{c}
x_+(t;\alpha)\\x_-(t;\alpha)
\end{array}
\right)
\]
or
\[
\left(
\begin{array}{c}
x_+(t+1;\alpha)\\x_-(t+1;\alpha)
\end{array}
\right)=
A_-
\left(\begin{array}{cc}
{m_+}^{\alpha}&0\\
0&{m_-}^{\alpha}
\end{array}
\right)
\left(
\begin{array}{c}
x_+(t;\alpha)\\x_-(t;\alpha)
\end{array}
\right)
\]
with probability $1/2$.
If $r_{\alpha}(t)$ is given as
\[
r_{\alpha}(t)=\frac{x_+(t;\alpha)}{x_-(t;\alpha)},
\]
then $r_{\alpha}(t)$ follows a one-dimensional Markov chain:
\[
r_{\alpha}(t+1)=\frac{(1-qu){m_+}^{\alpha}r_{\alpha}(t)+pu\ {m_-}^{\alpha}}
{qu\ {m_+}^{\alpha}r_{\alpha}(t)+(1-pu){m_-}^{\alpha}}
\]
or 
\[
r_{\alpha}(t+1)=\frac{(1-pu){m_+}^{\alpha}r_{\alpha}(t)+qu\ {m_-}^{\alpha}}
{pu\ {m_+}^{\alpha}r_{\alpha}(t)+(1-qu){m_-}^{\alpha}}
\]
with probability $1/2$.
Thus, the statistical state of $r_{\alpha}(t)$
is characterized by its stationary density distribution
$\rho_{\alpha} (r)$,
which can be numerically calculated \cite{lasota,morita02,morita12}.
In addition, when $x_+(t;\alpha)=r x_-(t;\alpha)$, we have 
\[
\langle x_i(t+1)^{\alpha}\rangle \simeq 
\frac{{m_+}^{\alpha}r+{m_-}^{\alpha}}{r+1}
\langle x_i(t)^{\alpha}\rangle.
\]
Consequently, 
as the exponent $\gamma$ is the critical value of $\alpha$, 
it is obtained by solving
\[
\int_0^{\infty}
\ln \left(\frac{{m_+}^{\alpha}r+{m_-}^{\alpha}}{r+1}\right) \rho_{\alpha}(r)dr=0.
\]
Setting $\alpha=1$ and using
the graphical method, 
we can obtain $\gamma$ for general values of $m_+$ and $m_-$.
Representative results are plotted in Fig.~3. Clearly, $\gamma$ increases as
$c$ increases or as $\tau$ decreases.

Model 2 assumes that the global environment
is auto-correlated as $\exp(-t/\tau)$. 
To obtain such a time series, we again apply the Markov chain with (\ref{eq5}).
At the same time step, the correlation between 
the multiplier $m_i(t)$
and the global environment is $1-2p$.
Thus, we obtain 
\begin{equation}
\langle (m_i(t)-\mu)(m_j(t')-\mu) \rangle
=\sigma^2 c \ e^{-|t-t'|/\tau},
\label{eq10}
\end{equation}
where
\[
c=(1-2p)^2.
\]
In this model, eq.~(\ref{eq10}) also holds for $i=j$. 
This condition marks an important difference between Models 1 and 2.
Because we have
\[
\rho_{\alpha}(r)=\frac{1}{2}\left[\delta(r-p/q)+\delta(r-q/p)\right],
\]
the exponent $\gamma$ is again given by eq.~(\ref{eq8}).
Thus, in Model 2, the exponent $\gamma$ is independent of the 
correlation time $\tau$.

To confirm the above predictions, numerical simulations are performed in (see Fig.~3).
To numerically estimate $\gamma$,
we set $N=10^6$,
employ the method of \cite{clauset}, and average over 200 samples.
Figure 3 shows that the theoretical predictions are highly consistent with the numerical simulation results.

In summary,
we investigated how the
spatio-temporal correlation of the multipliers
influences the power-law exponent.
On separately considering the temporal and spatial correlations,
we found that
$\gamma$ increased when
$c$ increased or when $\tau$ decreased.
In socioeconomic dynamics, such trends imply that the gap between the rich and the poor widens as the temporal correlation becomes stronger or as the
spatial correlation becomes weaker.
However, in a simultaneous treatment of the spatio-temporal correlation, the temporal correlation did not necessarily reduce the exponent (in Model 2).
At first sight, eqs.~(\ref{eq9}) and (\ref{eq10}) are similar and 
there is barely any difference between the correlation structures of Models 1 and 2.
However, this slight difference significantly affects the power-law exponent.

\begin{acknowledgments}
This work was supported by a Grant-in-Aid for Scientific Research (No. 26400388) and CREST, JST.
Some of the numerical calculations were performed on machines at YITP of Kyoto University.
\end{acknowledgments}


\begin{thebibliography}{99}
\bibitem{newman05}
M. E. J. Newman,
Contemp. Phys. {\bf 46}, 323 (2005). 

\bibitem{clauset}
A. Clauset, C. R. Shalizi and M. E. J. Newman,
SIAM Rev. {\bf 51}, 661 (2009).

\bibitem{gabaix09}
X. Gabaix,
Annu. Rev. Econ. {\bf 1}, 255 (2009).

\bibitem{gabaix99}
X. Gabaix
Q. J. Econ. {\bf 114}, 739 (1999).

\bibitem{ioannides}
Y. M. Ioannides and H. G. Overman,
Reg. Sci. Urban Econ. {\bf 33}, 127 (2003).

\bibitem{ramsden}
J. J. Ramsden and Gy. Kiss-Hayp\'{a}l
Physica A {\bf 277}, 220 (2000).

\bibitem{axtell}
R. L. Axtell,
Science, {\bf 293}, 18 (2001). 

\bibitem{mandelbrot}
B. Mandelbrot,
J. Bus. {\bf 36}, 394 (1963).

\bibitem{gabaix03}
X. Gabaix, P. Gopikrishnan, V. Plerou and H. E. Stanley,
Nature {\bf 423}, 267 (2003).

\bibitem{champernowne} 
D. G. Champernowne,
Econ. J. {\bf 63}, 318 (1953).

\bibitem{reed}
W. J. Reed,
Physica A {\bf 319}, 469 (2003).

\bibitem{pareto}
V. Pareto,
{\it Cours d'\'{e}conomie politique}.
F. Rouge, Lausanne (1896).

\bibitem{zipf}
G. K. Zipf, {\it Human Behavior and the Principle of Least Effort}. 
Addison-Wesley (1949).

\bibitem{render}
S. Render,
Am. J. Phys. {\bf 58}, 267 (1990).

\bibitem{sornette}
D. Sornette and R. Cont,
J. Phys. I {\bf 7}, 431 (1997).

\bibitem{biham}
O. Biham, O. Malcai, M. Levy, and S. Solomon,
Phys. Rev. E {\bf 58}, 1352 (1998).
 
\bibitem{sornette98}
D. Sornette,
Phys. Rev. E {\bf 57}, 4811 (1998).

\bibitem{nakao}
H. Nakao,
Phys. Rev. E {\bf 58}, 1591 (1998).

\bibitem{sato}
A. -H. Sato, H. Takayasu, and Y. Sawada
Phys. Rev. E {\bf 61}, 1081 (2000).

\bibitem{kesten}
H. Kesten, 
Acta. Math. {\bf 131}, 207 (1973).

\bibitem{morita13}
S. Morita and Jin Yoshimura, 
Phys. Rev. E {\bf 88}, 052809 (2013). 

\bibitem{lasota}
 A. Lasota and M. C. Mackey,
{\it Probabilistic Properties of Deterministic Systems}.
(Cambridge University, New York, 1985).

\bibitem{morita02} 
S. Morita and T. Chawanya,
Phys. Rev. E {\bf 65}, 046201 (2002).

\bibitem{morita12}
S. Morita and J. Yoshimura,
Phys. Rev. E {\bf86}, 045102R (2012).

\end{thebibliography}
\end{document}